\documentstyle[twocolumn,aps,epsf,floats]{revtex}
\begin{document}
\draft
\title{Simulations of interference effects in gated two-dimensional ballistic
electron systems}
\author{Antti-Pekka Jauho$^1$,
Konstantin N. Pichugin$^{2,3}$, and  Almas F. Sadreev$^{2,4}$}
\address{
$^1$Mikroelektronik Centret, Technical University of Denmark, Bldg 345east,
DK-2800 Lyngby, Denmark\\
$^2$Kirensky Institute of Physics, 660036 Krasnoyarsk, Russia 
\\
$^3$Institute of Physics, Academy of Sciences of the Czech Republic,
Cukrovarnicka 10, 16200 Prague, Czech Republic
\\
$^4$Dept. of Physics, \AA bo Akademi, SF-20500 Turku, Finland
\medskip\\
\date{\today }
\parbox{14cm}{\rm
We present detailed simulations addressing recent electronic interference
experiments, where a metallic gate is used to locally modify the Fermi
wave-length of the charge carriers. \ Our numerical calculations are based
on a solution of the one-particle Schr{\"o}dinger equation for a realistic
model of the actual sample geometry, including a Poisson equation based
determination of the potential due to the gate. The conductance is
determined with the multiprobe Landauer-B{\"u}ttiker formula, and \ in
general we find conductance vs. gate voltage characteristics which closely
resemble the experimental traces. \ A detailed examination based on quantum
mechanical streamlines suggests that the simple one-dimensional
semiclassical model often used to describe the experiments has only a
limited range of validity, and that certain 'unexpected' periodicities
should not be assigned any particular significance, they arise due to the
complicated multiple scattering processes occurring in certain sample
geometries.
\smallskip\\
Pacs numbers: 73.20.Dx,73.23.-b,73.23.Ad
\smallskip\\}}
\maketitle
\narrowtext
\section{Introduction}

Recent years have witnessed many experimental and theoretical advances
addressing the physical properties of mesoscopic samples, i.e., structures
where the phase of the electronic wave function directly affects the
measurable properties \cite{Meso}. \ A standard way to modify the phase of
the wave function is to use external magnetic fields: electrons traversing
the sample along a given path will accumulate a phase $\phi =\left(
e/h\right) \int {\bf A}\cdot d{\bf l}$, and thus give rise to interference
phenomena, such as Aharonov-Bohm oscillations. \ Recently Yacoby et al.\cite
{Yac1},\cite{Yac2} demonstrated another way of affecting the phase: a biased
metallic gate, placed above the two-dimensional electron gas, will change
the electron density (or, equivalently, the local Fermi wave-length)
underneath it,  and thus introduce a phase difference between electronic
paths that pass under the gate, and those that do not. \ In the first
experiment \cite{Yac1} the amplitude of the interference signal was used to
extract the energy, or temperature, dependence of the dephasing length in a
ballistic system. \ The experimental findings allowed a detailed comparison
with theoretical predictions \cite{eedphase} thus underscoring the
importance of this new technique. The second experiment \cite{Yac2} was the
first demonstration of a double-slit interference experiment in a solid
state system. \ Both of these experiments were analyzed with the help of the
following simple model. \ Assuming that the charge density is constant under
the gate (but different than elsewhere in the sample), Refs.\cite{Yac1},\cite
{Yac2} find that the phase difference $\Delta \phi $ of two representative
one-dimensional paths is given by $\Delta \phi =w\left( k_{F}-k_{F}^{\prime
}\right) =wk_{F}\left[ 1-\sqrt{1-\left( V_{g}/V_{{\rm dep}}\right) }\right] .
$ \ (Here $\ w$ is the width of the gate, $k_{F}$ and $k_{F}^{\prime }$ are
the Fermi momenta of the unmodified 2DEG and the 2DEG under the gate,
respectively, and $V_{{\rm dep}}$ is the gate voltage required to entirely
deplete the region under the gate \cite{comment}.) \ Indeed, the measured
conductances have a periodic component which essentially scales with the
square-root of the gate-voltage. \ In spite of this qualitative agreement,
some outstanding problems remain. \ In particular, a Fourier analysis of the
periodic signal of the double-slit experiment \cite{Yac2} contained an
unexpected low frequency component, approximately at half of the frequency
of the main feature. \ Yacoby et al. \cite{Yac2} tested a number of
plausible causes for this behavior (geometric effects, higher electronic
subbands, and spin-orbit interactions), but concluded that none of these
could satisfactorily explain the observations, which thus remained a puzzle.

The semiclassical picture discussed above is very persuasive, and indeed it
has been used in a large number of other contexts as well. \ The geometries
studied by Yacoby et al. \cite{Yac1},\cite{Yac2} are quite complicated, and
the possibility remains that an analysis based on one-dimensional straight
paths may miss some essential features. \ To the best of our knowledge,
these structures have not been analyzed in terms of a full solution of the
appropriate Schr\"{o}dinger equation, and the purpose of this paper is to
report such a study, the aim being the resolution of the problems
encountered in the interpretation of the second experiment \cite{Yac2}. \
Our work can be seen as a natural extension of several recent works
reporting detailed solutions of the Schr\"{o}dinger equation for
experimentally relevant semiconductor nanostructures \cite{literature}. The
resulting wave-functions often display a very rich structure and even
surprising physics, and as an example we mention vortices around nodal
points \cite{AlmasPRL}. \ The new ingredient in our work is that, in
addition to focusing on two recent experimental geometries, we include the
effect of the phase-modulating gate (PMG) on the potential landscape in
which the electrons move. \ Further, we
generate stream-lines from the probability-current flow; this allows us to quantify
the role played by the various paths contributing to conductance. \ Our main
conclusions are as follows. \ The effects of the gate can be felt in large
regions in the sample, and, in particular, in the double-slit geometry the
PMG also affects the slit region. \ It turns out that a description based on
a few characteristic paths works reasonably well in the geometry of the
earlier experiment \cite{Yac1}; this is not at all the case for the double
slit geometry. \ In general we find that the resulting conductance vs. gate
voltage curves are very sensitive to the details of the geometry. In
particular, the shape of the emitter and collector quantum point contacts is
found to play an important role. \ For certain parameters the simulated
conductance curves resemble closely their experimental counterparts, yet in
other cases, with nominally small changes in the parameter values, even the
qualitative appearance can change drastically. \ The numerical evidence
suggests that one should not assign major importance to specific features in
the Fourier transforms of the periodic conductance curves: they may just
reflect some details of the sample geometry and do not allow a simple
semiclassical interpretation.

We should note an important limitation of our numerical calculations. \ If
we express all energies in terms a parameter $E_{0}\equiv \hbar
^{2}/(2m^{\ast }d^{2}),$ where $d$ is the width of the injecting electrode,
the experiments typically involve energies of the order of 20000-30000 (we
estimate $d$ from published electron micrographs). \ Our numerical resources
do not allow energies much higher than 5000, i.e., one-fourth to one-sixth
of the experiments. \ As a consequence, the range of gate-voltages we can
study is somewhat smaller than what can be achieved experimentally, but
nevertheless we believe that our simulations have direct relevance on the
reported measurements. \ 

This paper is organized as follows. \ In Section \label{model}II we describe
the method of calculation, Section III is devoted to the analysis of the
first experiment, and Section IV presents our results for the double-slit
geometry.

\section{The Model}
\begin{figure}[tbp]
\epsfxsize=8.0cm
\epsfbox{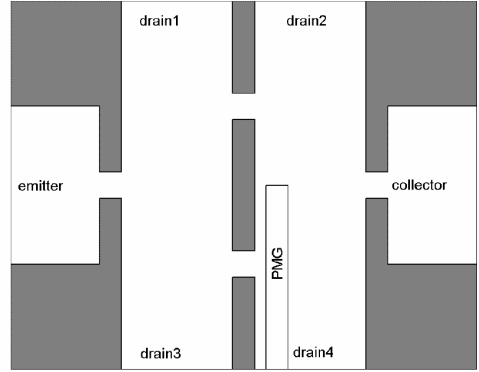}
\caption{Schematic representation of the six-terminal structure. The
boundaries of the structure and the double slit are shown as shaded areas;
they are modeled as hard walls. The lithographic placement for the
phase-modulating metallic gate (PMG) which is deposited on the surface of
the sample a height $h$ above the 2DEG (typically $h\simeq 100$\AA) is
indicated as a white rectangular area. The corresponding potential is shown
in Fig. 2.}
\label{fig1}
\end{figure}

The generic structure considered in this work is shown in Fig.\ \ref{fig1}.
\ It consists of the emitter, collector, phase-modulating gate (PMG), and
(possibly) the double slit (DS). \ The emitter and collector are modeled as
quantum point contacts. \ An important role is played by the four base
contacts: electrons scattering off from the DS or PMG, and not making it to
the collector leave the device via these contacts and do not contribute to
further interference patterns. \ In the simulations these ideal base
contacts are represented by open (Neumann) boundary conditions. In effect,
then, we are considering a six-terminal geometry (emitter, collector, and
four base contacts). \ The boundaries defining the structure are taken as
hard walls, and the potential describing the PMG is described below. Our
numerical method for the solution of the Schr\"{o}dinger equation is quite
standard, and here we only give those special features that are necessary
for understanding the computed results, given in several figures below.\ The
incident wave function in the $n$-th transverse channel is given by 
\begin{eqnarray}
\psi _{in,n}\left( x,y\right) &=&\sin \left( \pi nx\right)
e^{-ik_{n}y}\nonumber\\
&\quad&+\sum_{m}r_{nm}\sin \left( \pi mx\right) e^{ik_{m}y}
\label{psiin}
\end{eqnarray}
where the sine-functions are the transverse eigenfunctions of the injecting
electrode of unit width $\left( d=1\right) $, $r_{nm}$ are the corresponding
reflection coefficients, and the wave-vectors $k_{n}$ are defined via 
\begin{equation}
\epsilon =k_{n}^{2}+\pi ^{2}n^{2}  \label{kn}
\end{equation}
where the energy $\epsilon $ is given in units of $E_{0}$ defined above, and 
$k_{n}$ is in units of $1/d$. \ Analogously, the collector wave-function is
expressed as 
\begin{equation}
\psi _{out,n}\left( x,y\right) =\sum_{m}t_{nm}\sin \left( \pi mx\right)
e^{-ik_{m}y},  \label{psiout}
\end{equation}
where $t_{nm}$ is the transmission coefficient from mode $n$ to mode $m$,
and the collector is assumed to have the same width as the emitter. \
Finally, the boundary conditions at the $s$-th base contact (of width $L_{s}$%
) are specified by 
\begin{equation}
\psi _{s,n}\left( x,y\right) =\sum_{m}t_{s,nm}\frac{1}{\sqrt{L_{s}}}\sin
\left( \pi m\frac{y-y_{s}}{L_{s}}\right) e^{ik_{s,m}\left( x-x_{s}\right) }.
\label{psibase}
\end{equation}
Here $x_{s},y_{s}$ are the coordinates of the walls defining the base
contact $s$, and $k_{s,m}^{2}=\epsilon -\pi ^{2}m^{2}/L_{s}^{2}.$

The Hamiltonian in a tight-binding representation for the two-dimensional
electron system is 
\begin{equation}
\widehat{H}=-t(c_{i,j}^{{\dagger }}c_{i,j+1}+c_{i,j}^{{\dagger }}c_{i+1,j}+%
{\rm h.c.})+V_{i,j}c_{i,j}^{{\dagger }}c_{i,j}  \label{tbham}
\end{equation}
where $V_{ij}$ is the electrostatic potential due to the PMG. We use the
transfer matrix method as formulated in Ref.\cite{Ando} to compute the
various transmission and reflection coefficients (and hence the
conductances) \cite{computer}. \ 

\begin{figure}[tbp]
\epsfxsize=8.0cm
\epsfbox{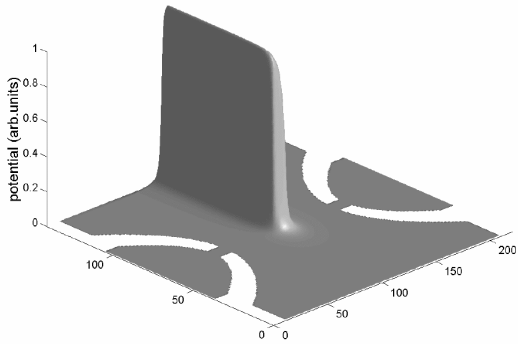}
\caption{The characteristic shape of the electrostatic potential due to the
phase-modulating gate. Also are shown curved emitter and collector contacts
as used in the experiment of Ref.[2].}
\label{fig2}
\end{figure}

A compact theory for the electrostatic potential caused by metallic gates on
the surface of the heterostructure has been developed by Davies et al.\cite
{Davies} for a number of different physical assumptions (pinned surface,
frozen surface, linear screening etc.). Following their analysis we choose 
\cite{OurChoice} (here $h$ is the distance between the gate and the 2DEG): 
\begin{eqnarray}
V\left( x,y,z=h\right) &=&\frac{V_{g}}{\pi } \bigl[ f\left( x,y-w/2,h\right)
\nonumber\\
&\quad&-f\left( x,y+w/2,h\right) \bigr] ,  \label{Vxyz}
\end{eqnarray}
where 
\begin{eqnarray}
f\left( x,y,h\right) &=&\arctan \left[ \frac{h}{R-x-y}\right] ,  \label{f} \\
R &=&\sqrt{x^{2}+y^{2}+h^{2}}.  \nonumber
\end{eqnarray}
A typical potential profile is shown in Fig.\ \ref{fig2}.

\section{The effect of phase-modulating gate}
\begin{figure}[tbp]
\epsfxsize=8.0cm
\epsfbox{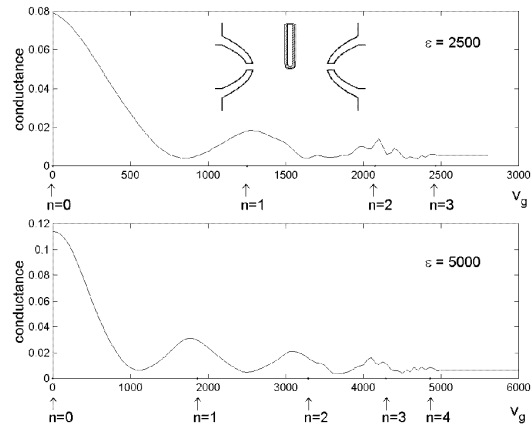}
\caption{Conductance vs. gate voltage for the device of Fig. 2 (shown as
inset) for initial energies $\protect\epsilon =2500$ (top) and $\protect%
\epsilon =5000$ (bottom). The predictions of the semiclassical formula for
conductance maxima are shown as arrows.}
\label{fig3}
\end{figure}

We first consider the experiment reported in \cite{Yac1}. \ The aim there
was to study dephasing due to electron-electron collisions in a ballistic
sample, and the phase modulating gate was introduced to generate an
interference signal: the amplitude of the oscillatory component of the
conductance is a direct measure of the phase-coherent part of the electrons,
and thus allows one to extract the phase-breaking rate as a function of
injection energy or temperature, and compare it to theoretical predictions 
\cite{eedphase}. \ The actual sample had two 'semi-infinite' gates, however
only one of them was activated and we therefore model this system with the
geometry shown in Fig. \ref{fig2}. Let us first construct an analytic
estimate for the expected behavior. \ Fixing the coordinates so that the PMG
runs parallel to the $x$-axis, and that the electron moves along the $y$%
-axis, the semiclassical formula for the phase accumulated under the PMG is 
\begin{equation}
\theta =\frac{1}{\hbar }\int dy\sqrt{2m^{\ast }\left( E-V\left( x,y,h\right)
\right) }.  \label{semiclphase}
\end{equation}
Neglecting the paths that pass near the edge of the PMG we can take the $%
x\rightarrow \infty $ limit of Eq.(\ref{Vxyz}), and find 
\[
V\left( y,h\right) =\frac{V_{g}}{\pi }\left[ \arctan \left( \frac{y+w/2}{h}%
\right) -\arctan \left( \frac{w/2-y}{h}\right) \right] 
\]
which, to a good approximation, provided that $h\ll w$, can be approximated
by a rectangular barrier of width $w$ and height $V_{g}$. \ The phase
difference between \ electrons passing under the PMG and those which do not
is then readily found to be
\begin{equation}
\Delta \theta =w\left( \sqrt{\epsilon }-\sqrt{\epsilon -v_{g}}\right) .
\label{ourphase}
\end{equation}
(Again, all energies are in units of $E_{0}$.) This derivation provides a
justification for the phenomenological expression for phase $\Delta \phi $
used in the experimental articles \cite{Yac1},\cite{Yac2}, see also \cite
{comment1}. \ Maximal constructive interference occurs when $\Delta \theta
=2n\pi $, which leads to expected maxima for PMG biases at 
\begin{equation}
v_{g}\left( n\right) =4\pi \sqrt{\epsilon }n/w-\left( 2\pi n/w\right) ^{2}.
\label{vgmax}
\end{equation}

\begin{figure}[tbp]
\epsfxsize=8.0cm
\epsfbox{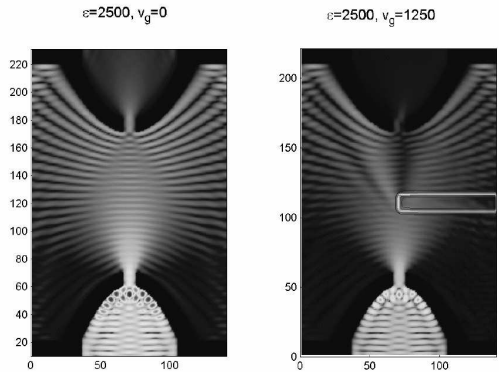}
\caption{The modulus of the wave-function for two gate voltages $V_g=0, 1250$
for incident electron in the first subband of the emitter with energy $%
\protect\epsilon=2500$.}
\label{fig4}
\end{figure}
Fig.\ \ref{fig3} shows the numerically computed conductance vs. PMG bias for
incident electron energies $\epsilon =2500$ (corresponding to 16 propagating
modes), and $\epsilon =5000$ (24 modes). \ The conductance was evaluated
with the Landauer-B\"{u}ttiker multichannel formula \cite{Buttiker},\cite
{Datta}: 
\begin{equation}
G_{EC}=\frac{2e^{2}}{h}T_{EC}=\frac{2e^{2}}{h}\sum_{nm}|t_{nm}|^{2}\frac{%
k_{m}}{k_{n}}.  \label{BLformula}
\end{equation}
The voltages at the base contacts are set to zero. \ The arrows in Fig.\ \ref
{fig3} indicate the predictions of the semiclassical formula, and one
observes a good overall agreement. As could be expected, the conductance
does not show any interference structure when the gate voltage becomes very
large, because the effective potential due to the gate is then
nontransmitting. \ 

In order to gain a deeper understanding of why the one-dimensional model
works so well in this particular case it is useful to study the quantum
mechanical stream-lines (for applications to several other physical systems
one may consult, e.g., \cite{Hirsch1}, \cite{WuSprung}, \cite{Lundberg}, 
\cite{Leavens}, \cite{Barker}). \ We use the following construction. \
Writing the wave-function in terms of an amplitude and a phase, 
\begin{equation}
\psi =\sqrt{\rho }\exp \left( iS/\hbar \right) ,
\end{equation}
the real and imaginary parts of the time-independent Schr\"{o}dinger
equation, 
\[
\left[ \frac{-\hbar ^{2}}{2m^{\ast }}\nabla ^{2}+V\right] \psi =E\psi ,
\]
yield 
\begin{eqnarray}
\frac{1}{2}m^{\ast }v^{2}+V+V_{QM} &=&E,  \label{EBohm} \\
{\bf \nabla }\cdot {\bf j} &=&0,
\end{eqnarray}
where 
\begin{eqnarray}
{\bf j} &=&\frac{1}{m}\rho {\bf \nabla} S,  \label{jBohm} \\
V_{QM} &=&-\left( \frac{\hbar ^{2}}{2m}\right) \left[ \frac{1}{2}\frac{%
\nabla ^{2}\rho }{\rho }-\frac{1}{4}\frac{\left( {\bf \nabla }\rho \right)
^{2}}{\rho ^{2}}\right] \nonumber\\
&=&-\frac{\hbar ^{2}}{2m}\frac{\nabla ^{2}\rho ^{1/2}}{%
\rho ^{1/2}}.  \label{VQM}
\end{eqnarray}
According to Bohm \cite{Bohm}, one interprets the electrons as ''real''
particles in the classical sense, following a continuous and causally
defined trajectory with a well-defined position ${\bf x}$ with the momentum
given by $m\stackrel{.}{{\bf x}}={\bf \nabla }S$. \ The force acting on the
particle is not derivable from the classical potential $V$ alone, but
acquires a quantum mechanical contribution from $V_{QM},$ Eq.(\ref{VQM}). \
The current stream-lines can then computed as in classical mechanics but
including the quantum force. The stream-lines can be viewed as an
alternative graphical presentation of the quantum mechanical probability
current density, see, e.g. Fig. 1 of Ref.\cite{AlmasPRL}. \ 
\begin{figure}[tbp]
\epsfxsize=8.0cm
\epsfbox{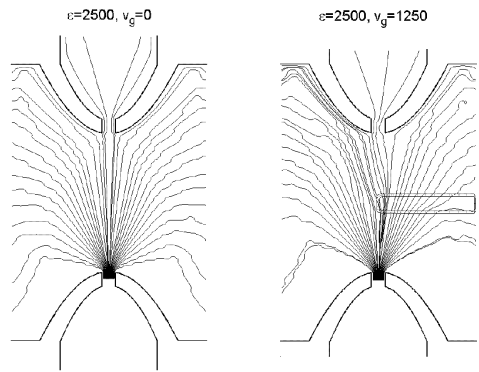}
\caption{Stream-lines for the gate voltages of Fig. 4. The stream-lines
passing under the gate and ending in the collector are essentially straight
lines and can be described by one-dimensional quantum mechanics.}
\label{fig5}
\end{figure}

Figure \ref{fig4} shows the computed wave-function for two different values
of the gate potential $V_{g}$, the right panel corresponding to the first
maximum in Fig. \ref{fig3}, top panel. \ We direct attention to the
following features. (i) The wave-functions display a rather regular pattern
even at a finite gate voltage (which breaks the mirror symmetry of the
problem). (ii) The curvilinear injector leads to a clear focusing effect. \
(iii) The effective wave-length is clearly longer under the gate than
elsewhere in the 2DEG, in accordance with the expectations. \ Fig. \ref{fig5}
shows the computed stream-lines. \ We note that most of the stream-lines
ending in the collector are, even in the case of a finite gate potential,
essentially straight lines. Thus they can be described by one-dimensional
quantum mechanics, and consequently the semiclassical estimate for
occurrence of conductance maxima, shown as arrows in Fig. \ref{fig3}, works
reasonably accurately. \ We next turn to the double-slit geometry, where
matters turn out to be quite different.

\section{Double slit geometry}

\begin{figure}[tbp]
\epsfxsize=8.0cm
\epsfbox{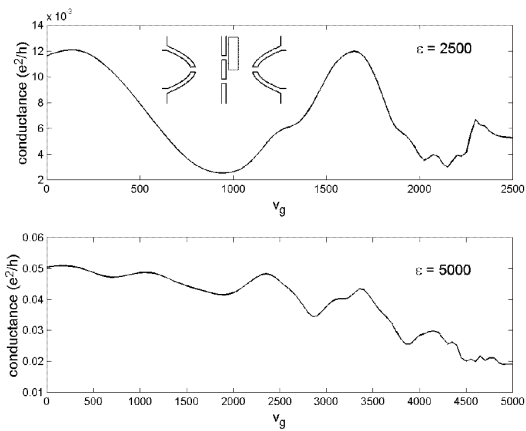}
\caption{Conductance vs. gate voltage for double-slit geometry, with curved
emitter and collector. The top view of the device is shown as an inset. }
\label{fig6}
\end{figure}
\begin{figure}[tbp]
\epsfxsize=8.0cm
\epsfbox{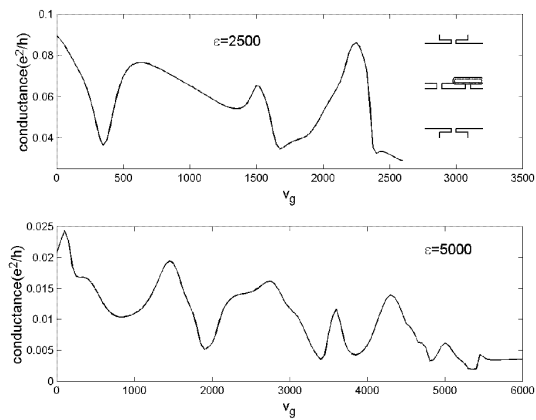}
\caption{Conductance vs. gate voltage for double-slit geometry, with
rectangular emitter and collector. A schematic view of the device is shown
as an inset.}
\label{fig7}
\end{figure}
We now introduce the double slit. \ In order to compare most directly with
the results obtained in the previous section, we first consider same
emitters and collectors as before, even though the experiment was done with
a different design (this will be discussed below). \ Fig. \ref{fig6} shows
the computed conductance in the presence of the double slit, as indicated by
the inset in the top panel. \ Again, we see oscillations in the conductance,
however the values of the gate voltage at which the conductance is at
maximum do {\it not} correspond to the values predicted by the simple
estimates, such as Eq.(\ref{ourphase}). We next consider the experimental
geometry of Ref. \cite{Yac2}, where the emitter and collector are
rectilinear. Figure \ref{fig7} shows the computed conductance curves and
\begin{figure}[tbp]
\epsfxsize=8.0cm
\epsfbox{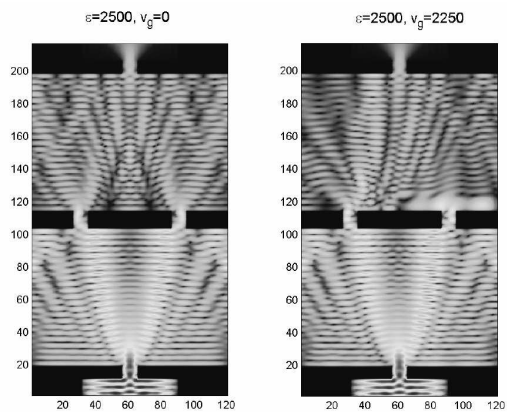}
\caption{The wave-function for double-slit geometry at two gate-voltages, $%
V_g=0$ and $V_g=2250$, the latter corresponding to the last significant
maximum in the conductance vs. gate-voltage characteristic of top panel of
Fig. 7.}
\label{fig8}
\end{figure}
Fig. \ref{fig8} displays examples of computed wave-functions. \ Conductance
oscillations are quite evident in Fig. \ref{fig7}, but it is much harder to
find any regular periodicity, in contrast to the curves for the device
without the double slit of Fig. \ref{fig3}. \ Since our simulated
conductance curves only have few maxima (because of the computational
restrictions to relatively low energies) we did not find a Fourier-analysis
helpful (as was the case for experiments which had a larger available gate
potential range): the resulting spectrum is dominated by spurious edge
effects. \ Finally, Figures \ref{fig9} and \ref{fig10} show the computed
stream-lines for the curved and rectangular emitters/collectors,
respectively. We draw attention to the qualitatively different picture as
compared to the device without the double-slit: the approximately
straight-line form of the stream-lines is almost entirely lost. Most
importantly, the stream-lines passing under the gate show a rather irregular
structure with \ a very wide range of effective path-lengths under the gate.
\ It is instructive to consider the pair of paths denoted by A and B (which
have symmetric initial conditions at the emitter): \ the combined effect of
the double-slit and the gate is to distort B significantly, and it is not
surprising that an one-dimensional model fails to describe it properly.
\begin{figure}[tbp]
\epsfxsize=8.0cm
\epsfbox{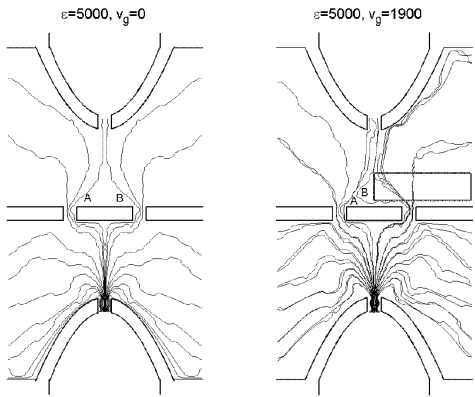}
\caption{Quantum stream-lines for double-slit geometry with curved emitter
and collector for gate voltages $V_g=0, 1900$.}
\label{fig9}
\end{figure}

The device with rectangular emitter/collector, as in \cite{Yac2}, has
another intriguing property: \ there is a clear tendency to form resonances
between the emitter and the double-slit (and, to a lesser extent, between
the double-slit and the collector). \ The effect of these resonances can be
understood in terms of a beating phenomenon: their frequency mixes with that
due to the PMG, and in general one can expect much more irregular
conductance vs. gate voltage curves, as is the case with curved
emitters/collectors. \ It is quite conceivable that this mixing can
contribute to the half-frequency oscillation observed in \cite{Yac2}.\
Another indication of these 'size-resonances' is that the conductance is not
always maximum at zero gate voltage (see, e.g., the lower panel of Fig. \ref
{fig7}): this is because the emitter quantum point contact is not always
matched to the resonator modes of the cavity formed by the gates defining
the emitter and the double-slit, and a finite gate-voltage can move the
resonator modes so as to achieve more efficient injection from the emitter.
\ In view of our simulations it would appear to be interesting to repeat the
double-slit experiment with curved emitter/collector: the experimental trace
is expected to be easier to interpret because one achieves a better focused
injection and diminishes complications due to the resonator modes.
\begin{figure}[tbp]
\epsfxsize=8.0cm
\epsfbox{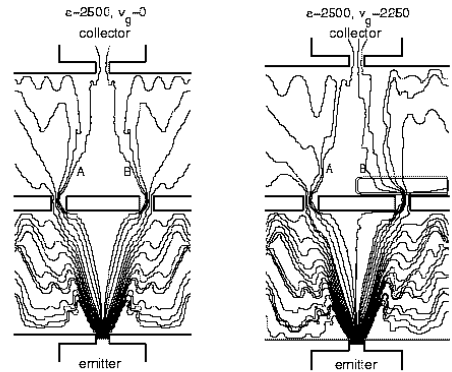}
\caption{Stream-lines for double-slit geometry with rectangular emitter and
collector.}
\label{fig10}
\end{figure}

\section{Conclusion}

We have presented simulations of phase-coherent charge transport in gated
mesoscopic structures. \ The simulations can describe the experiments at
least qualitatively, and under certain circumstances quantitatively. \ Using
quantum stream-lines as an interpretative tool we are able to offer an
explanation of why certain experiments can be interpreted with the help of
one-dimensional models, while others cannot. \ We find in general great
sensitivity to geometric effects, however these can be controlled at least
to some extent by careful device design aided with simulations of the kind
presented here, in particular when extended to higher energies.

\bigskip

\begin{acknowledgements}
This work was supported by the INTAS-RFBR Grant 95-IN-RU-657, RFFI Grant
97-02-16305, and the Krasnoyarsk Regional Science Foundation Grant 5F0060.
\end{acknowledgements}


\begin{references}
\bibitem{comment}  Assuming that the gate potential gives rise to an
effective square barrier of height $V_{{\rm eff}}\left( V_{g}\right) $, the
effective Fermi momentum under the gate is $k_{F}^{\prime }=k_{F}\sqrt{1-V_{%
{\rm eff}}\left( V_{g}\right) /\epsilon _{F}}$. \ The expression used in the
text is obtained by assuming that the effective barrier height scales
linearly with the gate potential, and reaches the Fermi energy at gate
voltage $V_{{\rm dep}}$: $V_{{\rm eff}}\left( V_{g}\right) =\epsilon
_{F}V_{g}/V_{{\rm dep}}$; $V_{{\rm dep}}$ can be extracted from experiment.
\ In Section III we show how the same expression emerges from a simplified
solution of the Laplace equation.
\bibitem{computer}  Our typical simulations have a mesh of $200\times 240$
sites, with a computational time of the order of 10 hrs.
\bibitem{OurChoice}  The potential for the semi-infinite phase-modulating
gate is constructed as a superposition of two infinite rectangles, see \cite
{Davies}, Eq. (3.9).
\bibitem{comment1}  It is, of course, quite straightforward to use the fully
quantum mechanical formula for the transmission coefficient over a square
barrier to compute the phase difference; a numerical test shows that the two
results are in very good agreement except for energies just above the
barrier. Thus the semiclassical formula used in the text is sufficient for
present purposes.
\end{references}
\end{document}